\documentclass{emulateapj}

\newcommand{\msun}{\ensuremath{\mathrm{M}_\odot}}

\def\gsim{\mathrel{\rlap{\lower 4pt \hbox{\hskip 1pt $\sim$}}\raise 1pt
\hbox {$>$}}}
\def\lsim{\mathrel{\rlap{\lower 4pt \hbox{\hskip 1pt $\sim$}}\raise 1pt
\hbox {$<$}}}

\begin{document}

\title{Prospect of Studying \\ 
Hard X- and Gamma-Rays from Type Ia Supernovae}

\author{
K.~Maeda\altaffilmark{1}, 
Y.~Terada\altaffilmark{2}, 
D.~Kasen\altaffilmark{3,4}, 
F.K.~R\"opke\altaffilmark{5,6},
A.~Bamba\altaffilmark{7}, 
R.~Diehl\altaffilmark{8}, 
K.~Nomoto\altaffilmark{1}, \\
M.~Kromer\altaffilmark{6}, 
I.R.~Seitenzahl\altaffilmark{5,6}, 
H.~Yamaguchi\altaffilmark{9}, 
T.~Tamagawa\altaffilmark{10,11}, 
W.~Hillebrandt\altaffilmark{6} 
}

\altaffiltext{1}{Kavli Institute for the Physics and Mathematics of the
  Universe (Kavli-IPMU), Todai Institutes for Advanced Study (TODIAS), 
University of Tokyo, 5-1-5 Kashiwanoha, Kashiwa,
  Chiba 277-8583, Japan; keiichi.maeda@ipmu.jp .}
\altaffiltext{2}{Department of Physics, Saitama University, 255
  Shimo-Okubo, Sakura-ku, Saitama 338-8570, Japan}
\altaffiltext{3}{Department of Physics, University of California at
  Berkeley, 366 LeConte, Berkeley, CA 94720, USA}
\altaffiltext{4}{Nuclear Science Division, Lawrence Berkeley National
  Laboratory, Berkeley, CA 94720, USA} \altaffiltext{5}{Universit\"at
  W\"urzburg, Emil-Fischer-Stra{\ss}e 31, 97074 W\"urzburg, Germany}
\altaffiltext{6}{Max-Planck-Institut f\"ur Astrophysik,
  Karl-Schwarzschild-Stra{\ss}e 1, 85741 Garching, Germany}
\altaffiltext{7}{Department of Physics and Mathematics, College of
  Science and Engineering, Aoyama Gakuin University, 5-10-1 Fuchinobe,
  Chuo-ku, Sagamihara, Kanagawa 252-5258, Japan}
\altaffiltext{8}{Max-Planck-Institut f\"ur Extraterrestrische Physik,
  Giessenbachstra{\ss}e 1, 85748 Garching, Germany}
\altaffiltext{9}{Harvard-Smithsonian Center for Astrophysics, 60
  Garden Street, Cambridge, MA 02138, USA} \altaffiltext{10}{RIKEN
  (Institute of Physical and Chemical Research), 2-1 Hirosawa, Wako,
  Saitama 351-0198, Japan} \altaffiltext{11}{Department of Physics,
  Tokyo University of Science, 1-3 Kagurazaka, Shinjuku-ku, Tokyo
  162-8601, Japan}

\begin{abstract}
We perform multi-dimensional, time-dependent radiation transfer
simulations for hard X-ray and $\gamma$-ray emissions, following
radioactive decays of $^{56}$Ni and $^{56}$Co, for two-dimensional
delayed detonation models of Type Ia supernovae (SNe~Ia). The
synthetic spectra and light curves are compared with the
sensitivities of current and future observatories for an exposure time of $10^6$ seconds. The non-detection of the $\gamma$-ray signal 
from SN 2011fe at 6.4 Mpc by SPI on board {\em INTEGRAL} places an upper limit for the mass of $^{56}$Ni of $\lesssim 1.0$ \msun, independently from observations in any other
wavelengths. Signals from the newly formed radioactive
species have not been convincingly measured yet from any SN~Ia, but 
the future X-ray and $\gamma$-ray missions are expected
to deepen the observable horizon to provide the high energy emission
data for a significant SN~Ia sample.  We predict that the hard X-ray
detectors on board {\em NuStar} (launched in 2012) or {\em ASTRO-H} (scheduled for launch in 2014) will reach to SNe~Ia at $\sim$15 Mpc, i.e.,
one SN every few years. Furthermore, according to the present
results, the soft $\gamma$-ray detector on board {\em ASTRO-H} will
be able to detect the 158 keV line emission up to $\sim$25 Mpc,
i.e., a few SNe Ia per year.  Proposed next generation $\gamma$-ray missions,
e.g., {\em GRIPS}, could reach to SNe Ia at $\sim$$20 - 35$ Mpc by MeV observations. Those would provide new diagnostics and strong constraints on explosion models, detecting rather directly the main energy source of supernova light. 
\end{abstract}


\section{Introduction}

It is widely accepted that Type Ia Supernovae (SNe Ia) are a major
source of Fe in the Universe \citep[see e.g.,][for a
review]{hillebrandt2000}.  The thermonuclear explosion of a white
dwarf produces $^{56}$Ni as a main product
\citep[e.g.,][]{nomoto1984}.  It decays into $^{56}$Co (with an e-folding time of $\sim$8.8 days) and then into $^{56}$Fe ($\sim$113
days). The transitions typically occur into excited states of the daughter nuclei, 
which generally de-excite by emissions of $\gamma$-rays with characteristic energies of $\sim$1 MeV. 
These $\gamma$-rays energize the thermal electron pool mainly through
Compton scattering, which ultimately leads to the optical appearance
of SNe Ia.

Although this scenario has been supported by studying optical spectra
and light curves, the most direct evidence is still missing, i.e., the
detection of the decay $\gamma$-rays and related high energy emissions
\citep[e.g.,][]{clayton1969,ambwani1988,milne2004}.  A marginal
detection for the peculiar SN Ia 1991T was reported
\citep{lichti1994,morris1997}, while for the more-nearby SN Ia 1998bu
only upper limits were obtained \citep{georgii2002}.  The recently
discovered SN Ia 2011fe in the nearby galaxy M101 at $\sim$6.4 Mpc
\citep{nugent2011a,nugent2011b} has been observed by {\em INTEGRAL}, but only an upper limit has been placed \citep{isern2011a,isern2011b}.

Despite the importance of multi-dimensional structures in 
state-of-the-art explosion models \citep[e.g.,][]{gamezo2003,
roepke2005,bravo2006,roepke2007,jordan2008,seitenzahl2011} up to now most theoretical 
studies of the high energy emission of SNe~Ia have been restricted to 
one-dimensional models \citep[e.g.,][]{ambwani1988,hoflich1992,
hoflich1998,gomez1998}. For a review of the (one-dimensional) 
theoretical studies of high energy signals from SNe~Ia see 
\citet{milne2004}. Only recently, first multi-dimensional studies
became available \citep[e.g.][]{hoflich2002,hungerford2003,maeda2006,
sim2008,kromer2010}. \citet{hoflich2002} discussed effects of the multi-dimensionality
in the explosion (for a few delayed detonation models) on the line profiles of the 812 and 847 keV lines. \citet{sim2008} examined flux
evolutions in different energy bands based on kinematic models. 
\citet{kromer2010} presented a prediction of the high energy signal based on the double-detonation sub-Chandrasekhar models. 
In this paper, we report the first study on expected high energy emission signatures and their flux evolutions from a series of two-dimensional delayed-detonation models. 

We use these models to discuss possible constraints on the explosion
mechanism of SNe~Ia through the $\gamma$-ray non-detection from SN 2011fe by the currently
operating instrument SPI on board {\em INTEGRAL} \citep[a narrow line
sensitivity of $3.1 \times 10^{-5}$ photons cm$^{-1}$ s$^{-1}$ at 1 MeV for
$10^{6}$ seconds exposure:][]{roques2003}.\footnote{We adopt the
  sensitivity of SPI from the latest SPI observer's manual at
  http://www.rssd.esa.int/ .} We then examine the detectability of the
high energy emission from SNe Ia by near future observatories.  Those
include {\em NuStar} \citep{koglin2005}, HXI \citep[Hard X-ray Imager:
][]{kokubun2010} and SGD \citep[Soft Gamma-ray Detector:
][]{tajima2010} on board {\em Astro-H} \citep{takahashi2010}, and {\em
  GRIPS} \citep{greiner2011}.  {\em NuStar} and HXI are designed to
reach to a sensitivity of a few $10^{-8}$ cm$^{-2}$ s$^{-1}$
keV$^{-1}$ (for $10^6$ seconds) in the hard X-ray range.  The SGD's
designed sensitivity is (5 -- 10) $\times 10^{-8}$ cm$^{-2}$ s$^{-1}$
keV$^{-1}$ (for $10^6$ seconds) in the soft $\gamma$-ray range below
$\sim$600 keV.  {\em GRIPS} is one example for proposed next
generation telescopes, designed to be by a factor of $\sim$15 better
in sensitivity than {\em INTEGRAL} in the MeV range.

\section{Methods and Models}
We have performed radiation transfer calculations for the series of
two-dimensional delayed detonation models of \citet{kasen2009}
\citep[see also][]{maeda2010a} (Fig. 1).  In these models, the
thermonuclear explosion was initiated in a number of randomly placed
sparks near the center of the white dwarf. The combustion front
initially propagates as a subsonic deflagration and later turns into a
supersonic detonation \citep{khokhlov1991}.  This scenario involves a
deflagration-to-detonation transition, which in our models is
parameterized in terms of the turbulence strength required at the
transition spot. Since the microphysics of deflagration-to-detonation transitions remains an open question, \citet{kasen2009} suggested a set of criteria that initialize the transition at different Karlovitz numbers. Here, we consider models employing two of these criteria, namely the sets dc2 and dc3. The latter requires a larger Karlovitz number and thus delays the transition. Consequently, the detonation proceeds at lower densities and the resulting explosion is typically fainter.
Predicted optical emission properties from these models are generally in good agreement with
observational properties of normal SNe Ia
\citep{kasen2009,maeda2010b}, while those with the largest asymmetry
have some problems \citep{maeda2010c,blondin2011}.  The simulations
were done for the C+O white dwarf at the solar metallicity. The
detailed nucleosynthesis was followed for a few selected models,
and the result was used to interpolate the elemental distribution for
C, O, Na, Mg, Si, S, Ar, Ca, Ti, Cr, Fe, Co, Ni in the other
models. Since the explosive nucleosynthesis is well characterized by
peak temperature and density at the passage of the flame, this
procedure would not introduce a large error in the abundance
pattern. For each case, 16 different distributions for the initial
thermonuclear sparks were investigated.  These are divided into two
sequences, `iso' and `asym' (named either DD2D\_iso\_AA or
DD2D\_asym\_AA, with AA spanning from 01 to 08).  In the `iso' models,
the ignition points were randomly distributed in a sphere around the
white dwarf center, while in the `asym' models the ignition points
were preferentially offset in a certain direction.  The resulting mass
of $^{56}$Ni ranges from 0.34 to 1.15 \msun (see Tab. 1 for
examples).  In the `iso' models, the number of ignited sparks is
larger for a larger value of AA. In the `asym' models, the ignited
sparks were put in a narrower cone for larger AA with a smaller number
of the sparks. Generally, a smaller number of initial sparks (smaller
AA in `iso' and larger AA in `asym') leads to a weaker deflagration
and a stronger detonation, resulting in larger $M$($^{56}$Ni).  We
have also investigated the one-dimensional pure-deflagration model W7
\citep{nomoto1984}.  In total, 33 models have been investigated.

Transfer of $\gamma$-rays from the decay chain $^{56}$Ni $\to$
$^{56}$Co $\to$ $^{56}$Fe has been followed using the
three-dimensional time-dependent radiation transfer code of
\citet{maeda2006} \citep[see also][]{kasen2006,sim2008}. In our
calculations, Compton scattering, pair creations, and photoelectric
absorptions have been included as interaction processes.  For Compton
scattering, the Klein-Nishina cross section is used. For pair production, cross sections are
adopted from \citet{hubbell1969}. For photoelectric absorption, cross
sections compiled by \citet{hoflich1992} from \citet{veigele1973} are
used, with the same interpolation scheme adopted by \citet{hoflich1992}. 
A test calculation for the transfer code using the W7 model
shows a good match to the spectral sequence obtained by a majority of
the transfer codes \citep[e.g.,][]{hungerford2003} presented in \citet{milne2004} 
over the energy range between 80 keV and 2 MeV for which
the synthetic spectra for the comparison are available in
\citet{milne2004}. We are thus confident that our prediction is solid
in the soft $\gamma$-ray energy. \citet{milne2004} noted that there are 
differences in the energy range below 100 keV depending on the detail
of the transfer schemes -- the variation in the predicted flux just
below 80keV is likely less than 50\% \citep[see Figs. 4 and 5
of][]{milne2004}, but one should keep in mind that this is the level of
uncertainty involved in the predicted hard X-ray flux. We assume 
direct annihilation for the 511 keV lines without positronium
formation. This will affect the 511 keV line flux to some extent, but
the effect of the positronium formation on the down-scattered
continuum is not important; the positronium decay continuum emission
is concentrated just below 511keV and the Compton scattering cross
section in this energy range is almost identical to that at 511
keV. Indeed, \citet{milne2004} found that there is very little
difference between the results from codes that include a
positronium continuum component and those that do not.

The models have been mapped onto $129^3$ spatial grids. In each
simulation, we follow $10^9$ photon packets with continuous
changes in the spatial position, direction, and spectral energy and
energy content. Once they escape from the ejecta they are binned into
10 angular bins, 3000 spectral energy bins, and 36 logarithmically
spaced time bins from 5 to 300 days after the explosion. This yields
time- and angle-dependent spectra, from which light curves are extracted for specific energy bands.

\section{Results}

\subsection{Spectra}

Figure 2 shows examples of synthetic spectra.  The flux is normalized to a distance of 10 Mpc.  Also shown are designed sensitivity curves
of HXI and SGD on board the planned {\em ASTRO-H}
\citep{kokubun2010,takahashi2010, tajima2010}.

The $\gamma$-ray lines are produced by radioactive decays (including
the 158 keV and 812 keV lines from $^{56}$Ni decay and the 847 keV
line from $^{56}$Co decay). The lines are Doppler shifted by
the bulk expansion. Initially only emissions from the outer, thin layer along the line-of-sight is  seen as a (narrow) blueshifted line. As time goes by, the escaping lines become broader and the line-center moves to lower energy, following the increasing transparency in the deeper, slower part of the ejecta (resulting in the thicker layer probed in the lines, thus the lines become broader). Compton scattering produces the continuum
below the line energies, with the low energy cut off created by
photoelectric absorptions. The cut off energy becomes higher as time
goes by, due to the increasing contribution to the emission from the
deeper part, where the mean atomic number and photoelectric cross
sections are larger.

At 20 days after the explosion, larger dependence on the viewing
direction is seen in DD2D\_iso\_04 than in DD2D\_asym\_04, despite the
initial large asymmetry in the ignition in the latter.  Generally,
fainter models with smaller $M$($^{56}$Ni) show a larger viewing-angle
dependence, and this is a specific model prediction for the asymmetric
delayed detonation models \citep{kasen2009,maeda2010a}.  Even if the
initial asymmetry is large, a strong detonation tends to decrease the
inhomogeneity in the final abundance distribution.  Faint models with
a weak detonation preserve the inhomogeneity in the distribution of
$^{56}$Ni either created by the initial asymmetry and/or mixing in the
deflagration phase \citep{kasen2009} (Fig. 1).

At 20 days, the 2D delayed detonation models show harder spectra than
the W7 model. This is a result of more extended burning and larger
photoelectric absorption cross sections near the surface in the 2D
models.  For example, in the W7 model the mass fraction of Si exceeds
0.1 up to 15,000 km s$^{-1}$ (with the inner boundary of $\sim$10,000
km s$^{-1}$) \citep{nomoto1984}, while in the DD2D\_iso\_04 model it
exceeds 0.1 up to $\sim$14,000 - 20,000 km s$^{-1}$ depending on
the viewing direction despite smaller $M$($^{56}$Ni) (Fig. 1).  Models
with a strong viewing-angle effect (i.e., faint models) show a harder
X-ray cut-off when observed from brighter line-of-sight.  The large
flux is created by a larger amount of $^{56}$Ni near the surface in
this direction.  This, together with other burning products like Si,
also increases the photoelectric absorption cross section in the same
direction.  At 60 days, the optical depth decreases and the viewing
angle effect becomes weak \citep[see also][]{sim2008}.

\subsection{Light Curves}

Figure 3 shows light curves for selected energy bands and
lines. Figure 4 shows examples of evolution in ratios between two
different bands. Figure 5 shows selected continuum and line fluxes for 
the reference distance 10 Mpc, for all the models and viewing
directions as a function of $M$($^{56}$Ni). Figure 6 is for a 
constraint on the nature of SN 2011fe -- the predicted flux in the energy range 
$830 - 875$ keV at 6.4 Mpc. The 158 keV line strength has been extracted from the
synthetic spectrum in the energy range $150 - 168$ keV. The 812 keV and
847 keV line strengths have been extracted as follows. First, we integrate
the synthetic spectra in the energy range $790 - 900$ keV, then the
total flux is divided into the two lines according to the decay
probabilities at each epoch (thus assuming the same cross section for
these two lines, which should be a good approximation).

The sensitivities of HXI and SGD are calculated by integrating the
designed sensitivity curves
\citep{kokubun2010,takahashi2010,tajima2010} in the corresponding
energy ranges. For lines, we have integrated the sensitivity curve
within the typical FWHM predicted by the models, which evolves from
$v/c \sim 0.02$ to $0.035$ from 20 to 80 days.  We note that for a
detailed comparison between the models and observations, differences
in the line width need to be taken into account --- because of the
larger line width and a larger background contamination in brighter
models, these models are harder to detect than the first-order
estimate provided here (the opposite is true for the fainter models).
For {\em NuStar}, we have assumed a constant sensitivity across the
energy band, $4$ (conservative) and $2$ (optimistic) $\times 10^{-8}$
cm$^{-2}$ s$^{-1}$ keV$^{-1}$ \citep{koglin2005}.  For SPI, the narrow
line sensitivity is multiplied by the degradation factor from the line
broadening, which we assume is $\sqrt{FWHM/\Delta E}$ \ with a typical
resolution $\Delta E = 2$ keV.  The degradation factor we adopt is
therefore $\sim$1 to 4, depending on the energy range and the
epoch. It would be
possible to reduce the resulting degradation when background features
are well constrained and stable, and/or when specific line shapes can
be assumed or tested.  For {\em GRIPS} we simply assume that the
narrow-line sensitivity is better than SPI by a factor of 15
\citep{greiner2011}.

In the hard X-ray continuum, the 2D delayed detonation models peak
fainter than the W7 model even if $M$($^{56}$Ni) is larger, as a
result of the larger mean atomic number (due to more complete burning)
and larger photoelectric absorption cross sections near the
surface. The signal therefore can be a strong diagnostic for the
composition near the surface and the mode of the flame propagation.
The models with larger $M$($^{56}$Ni) peak earlier, at a higher flux
level: thus the hard X-ray alone (as well as the 158 keV line
discussed below) could give a rough constraint on $M$($^{56}$Ni)
independently from optical observations.  The 2D models show a strong
angle-dependence. The angle variation in the model flux reaches nearly
100\% for models with $M$($^{56}$Ni) $\sim 0.6$ \msun, while 
for the brighter models it is at most $\sim$40\% (Fig. 5).  If an SN Ia
as close as SN 2011fe would appear again in the coming decade when the
near-future hard X-ray instruments (HXI and {\em NuStar}) will be
operating, the signal should be detectable by these instruments.

The behavior of the 158 keV line ($^{56}$Ni $\to$ $^{56}$Co) is
similar to that of the hard X-ray continuum (see above), except that
the difference between the 2D models and W7 is not large for given
$M$($^{56}$Ni), since the dominant interaction process is Compton
scattering and the difference in the surface composition is not
important.  Not many previous theoretical studies have explicitly
addressed the 158 keV line strength -- in Figure 3 we plot the
prediction on the peak 158 keV flux from \citet{gomez1998} together
with our 2D delayed detonation model predictions. The models of
\citet{gomez1998} are 1D models, including a delayed-detonation
model [$M$($^{56}$Ni) $= 0.8$ \msun], a pure-detonation model
[$M$($^{56}$Ni) $= 0.7$ \msun], and a sub-Chandrasekhar explosion
model [$M$($^{56}$Ni) $= 0.6$ \msun].  
In \citet{gomez1998} the 158 keV flux was extracted  from an energy range corresponding to 
$\sim 1.2 \times$ FWHM. Thus, their line fluxes were extracted from a narrower range than what we adopt for our models. The difference should not be large as we deal with a line. Their delayed-detonation model follows our (angle-averaged) model behavior. The pure-detonation
model predicts a larger peak flux than any of the 2D delayed-detonation
models, and the difference is about a factor of three for the same
$M$($^{56}$Ni). The sub-Chandrasekhar model peaks earlier than for the
delayed detonation model, and the peak flux is higher by
about a factor of two than the corresponding delayed detonation model
with similar $M$($^{56}$Ni).  All of our delayed detonation models, 
as well as other typical models \citep[e.g.,][]{gomez1998} are reachable 
by SGD for an SN at 10 Mpc (Fig. 3).

The 158 keV line is generally the strongest signal in the first 
month. Since the expected behavior is sensitive to the distribution of
$^{56}$Ni, it is most useful to distinguish models which result in
a very different spatial distribution of $^{56}$Ni. For SN 2011fe, the
$2\sigma$ upper limit on the integrated flux in the energy range $160-166$ keV at
$\sim$10 days after the explosion has been placed at $7.5 \times
10^{-5}$ photons cm$^{-2}$ s$^{-1}$ by the SPI observation with an
exposure of $\sim$$10^{6}$ second \citep{isern2011a}. This upper limit
rejects none of our models, and neither of the other models shown in
Figure 3 -- our model fluxes extracted in the energy range $150-168$ keV are lower than 
$2 \times 10^{-5}$ photons cm$^{-2}$ s$^{-1}$, and should be even lower for the $160 - 166$ keV range. 
The pure-detonation model in \citet{gomez1998} has the
largest flux among the models presented here, $\sim$$6 \times 10^{-5}$
photons cm$^{-2}$ s$^{-1}$ at the distance of 6.4 Mpc which is still
below the observational upper limit.

The light curve of the $200 - 460$ keV range and that of the 812
keV line follow a behavior similar to that of the 158 keV line (see
below for more detailed discussion on the $200-460$ keV continuum).
In Figures 3 and 5, we show the prediction of other well-investigated
models together with our 2D delayed-detonation models, for the 812 keV line 
flux. These model
predictions have been taken from \citet{milne2004}: Normal-brightness
SN Ia models (HED8, DD202c, W7), luminous models (HECD, W7DT),
and sub-luminous models (HED6, PDD54) covering a wide range of
$M$($^{56}$Ni), $0.14 - 0.76$ \msun. The models which have a large
amount of $^{56}$Ni near the surface peak earlier and are brighter than
our 2D delayed-detonation model sequence for given $M$($^{56}$Ni). It
is seen that the typical behavior of the 2D delayed-detonation model
is similar to the 1D deflagration model W7 and the 1D delayed
detonation model DD202c, except that a variation in the flux is
expected for different viewing directions in the 2D models (like in
the hard X-ray continuum and the 158 keV line). We note that \citet{milne2004} integrated the flux for the 812 and 847 keV lines 
in the energy range $810 - 885$ keV, narrower than our definition ($790 - 900$ keV). However, the continuum level is quite low, thus the difference is negligible (i.e., at most a few per cent). 

The 847 keV line ($^{56}$Co $\to$ $^{56}$Fe) is the strongest line for SNe Ia and it peaks around 
$60 - 100$ days after the explosion. This
line is insensitive in its behavior to the viewing direction due to a
small optical depth in these late epochs. Indeed, the behavior of this
line is rather insensitive to model variants as well, but mostly
determined by $M$($^{56}$Ni), up to $\sim$30\% level, as shown in Figure 5. This means
that the detection of this signal around the peak would not discriminate different
model variants (such as pure-detonation, delayed detonation, 1D or
multi-D), but provides a direct estimate of $M$($^{56}$Ni) where the
uncertainty would come mostly from the limited measurement accuracy rather than
from the model uncertainties.

A constraint from the reported upper limit on the $\gamma$-ray signal
from SN 2011fe by SPI/{\em INTEGRAL} \citep{isern2011b} is the
following. With $\sim$$3 \times 10^{6}$ second exposure at days $45 - 88$,
they placed a preliminary 95\% upper limit (2$\sigma$) of $1.39 \times 10^{-4}$ photons cm$^{-2}$
s$^{-1}$ in the energy range $830 - 875$ keV and $1.20 \times 10^{-4}$
photons cm$^{-2}$ s$^{-1}$ in $835 - 870$ keV. Adopting typical model predictions for the line profile, the former likely includes most of the
flux in the 847 keV line, while the latter will miss about $10 - 15$\%
of the flux. Thus, the constraints from these different band passes
are similar. In Figure 6, we show our model predictions at 6.4
Mpc, where the model fluxes are averaged over the SPI observation time window ($45 - 88$ days). The reported upper limit rejects the models with $M$($^{56}$Ni)
$\gsim 1.0$ \msun. 

The constraint here should be compared with the optical properties of
SN 2011fe. The optical light curve evolution suggests that it is a
normal SN Ia peaking at $-19.13$ mag in $B$ band, indicating that
$0.45\ \msun \lsim $ $M$($^{56}$Ni) $\lsim 0.6\ \msun$ have been produced in the explosion
\citep{nugent2011b,roepke2012}. \citet{roepke2012} presented
optical-spectral analysis based on two explosion models, a 3D delayed
detonation model (similar to those presented in this paper) and a
white dwarf-white dwarf merger, both models having $M$($^{56}$Ni)
$\sim 0.6\ \msun$. Both models show a general agreement with the
observed spectral sequence: \citet{roepke2012} found a slightly 
better match for the merger model but could not place a clear preference 
based on the available optical data. The constraint from the SPI observation,
$M$($^{56}$Ni) $\lsim 1.0$ \msun, is consistent with the
optical emission analysis, but unfortunately not deep enough to
provide further constraints. However, we emphasize that the constraint
from the high energy emission is fully independent of the optical
emission analysis and suffers from much less theoretical
uncertainties. Thus, it demonstrates that a strong constraint can
potentially be obtained from high energy signals for future nearby SNe
Ia.

We predict different behaviors for different energy ranges, depending
on the models and viewing directions (see above). This provides strong
diagnostics on the explosion mechanism.  Figure 4 shows two examples;
the flux ratio between the 158 keV line and the $200 - 460$ keV
continuum, and that between the $60 - 80$ keV continuum and $200 - 460$ keV
continuum. The first example is sensitive to the column mass density above the 
$^{56}$Ni-rich region. The dominant interaction process at 158 keV
and in the high energy continuum is Compton down-scattering (there is
significant contribution from the photoelectric absorption at 158 keV,
but this contribution is negligible in the argument). The strength of
the 158 keV line is proportional to the escape fraction of photons,
while that of the $200 - 460$ keV continuum is the combination of the
scattering fraction of the decay lines and the escape fraction of the
down-scattered photons. As such, the ratio is larger for smaller optical depth due to Compton scattering (i.e., for smaller amount of material above $^{56}$Ni). It is seen in Figure 4 that models with
large amount of $^{56}$Ni and/or those viewed on-axis have a high
flux ratio (i.e., a situation in which the distribution of
$^{56}$Ni is extended toward the surface along the line-of-sight). The
ratio between the hard X-ray continuum and the soft gamma-ray
continuum behaves in the opposite way. This is an example of the
diagnostics about the chemical composition near the surface. At
$60 - 80$ keV, the photoelectric absorption is a dominant interaction
process. As such, the ratio here is a measure of the average atomic
number near the surface through the cross section to the photoelectric
absorption. Models with the largest abundance of Fe-peak elements
near the (line-of-sight) surface have the smallest ratio.

\subsection{Detectability}

Table 1 summarizes the expected detectability of a few selected 
models. The exposure of $10^{6}$ seconds is assumed to be centered on the peak flux date. The time-averaged fluxes extracted from the spectral time series are then compared with sensitivities of various instruments. In practice, the peak date is not known in advance. This is not a big problem for the 847 keV line, but it matters for the low energy bands, e.g., the hard X-rays and the 158 keV line, where the peak date is sensitive to the model details (see Fig. 3). In planning observations, it is thus encouraged to take into account the evolution of the optical emission so that the coordinated observation covers the peak flux. This, however, requires a very early discovery and intensive follow-up as was done for SN 2011fe \citep{nugent2011b}. Table 1 also gives the expected number of SNe Ia per year based on the Asiago Supernova Catalog that is constantly updated with new SN discoveries \citep{barbon1999} \citep[see][for the local SN Ia rate]{horiuchi2010}. 

In the well-studied 847 keV line, we estimate that SPI on board {\em INTEGRAL}
reaches to SNe of normal brightness at $\sim$6 Mpc, in an ideal situation. 
This can be optimistic though, as in practice the SPI observation of SN 2011fe with $3 \times 10^{6}$ seconds did not detect any positive signals for the data analysis performed so
far. In our estimate of the SPI detectability, we have not included the sensitivity loss from modeling the background that could further reduce the sensitivity beyond the increased statistical fluctuations when using a broad energy band. This loss of the sensitivity depends on the quality of the background modeling in the data analysis, so further analysis of the SPI data would reduce the upper limit below the reported value by \citet{isern2011a,isern2011b}. The observation (with the data analysis so far) does constrain $M$($^{56}$Ni) $\lsim 1.0$ \msun produced in SN 2011fe.  A future telescope like {\em GRIPS}
is estimated to reach to $\sim$20 Mpc in the 847 keV line, if the
sensitivity is improved by a factor of 15 compared to {\em SPI} as is
designed. Since the 847 keV line luminosity evolves slowly around the peak, even an
exposure as long as $3 \times 10^{6}$ seconds could be coordinated \citep[as
was done for SN 2011fe:][]{isern2011b}, and then the detection horizon
will even extend to $\sim$35 Mpc.

We have found that the near-future instruments for hard X-rays and
soft $\gamma$-rays, {\em NuStar} (launched in July 2012) and {\em
  ASTRO-H} (launch planned for 2014), will potentially provide
detection of the high energy emission from extragalactic SNe Ia almost
annually. {\em NuStar} and HXI are estimated to reach to SNe Ia up to
$\sim$15 Mpc (for SNe Ia of normal luminosity), i.e., roughly one SN
every few years. The 158 keV line will be detectable by SGD up to a
distance of $\sim$25 Mpc, i.e., a few SNe per year. The continuum
between 200 and 460 keV will be detectable up to $\sim$14 Mpc.  Even
if we make a very conservative assumption that the SGD sensitivity
curve would be degraded from the present design by a factor of three, 
normal SNe Ia at $\sim$15 Mpc will be reachable (Tab. 1). We note
that the model prediction is more solid for the 158 keV line than for
the hard X-rays; the degree of the agreement between different
transfer codes (tested for the same model) is worse in the hard X-rays
due to more complicated cross sections in this energy range
\citep{milne2004} (see \S 2).  The hard X-ray and the 158 keV line
evolve relatively quickly, but the flux levels stay almost constant around peak
for $\sim$$10 - 20$ days, thus an exposure of $10^6$ seconds is
appropriate.  It also requires an SN discovered soon after the
explosion. Such an early detection is now feasible thanks to new
transient searches with high cadence \citep[e.g.,
][]{nugent2011a,nugent2011b}.

Another possible concern on the hard X-ray band study is a
contamination from background sources.  The predicted peak hard X-ray
luminosity from an SN is $\sim$$10^{39}$ erg s$^{-1}$.  For example, in M101 there
are $\sim$100 X-ray sources above $\sim 10^{36}$ erg s$^{-1}$ in the
energy range 0.1 -- 8 keV \citep{pence2001}. For a conservative
angular resolution of 2 arc minutes for HXI, about 5 or 10 such
sources may be unresolved from the SN emission. The expected total
flux from these contaminated sources is however far below the
predicted SN hard X-ray luminosity.  A probability that a host galaxy
has a powerful active galactic nucleus and an SN is close to the core
is small for nearby SNe (neither is the case for SN 2011fe).  Also,
the characteristic temporal evolution of the SNe, i.e., the quick
decrease of the flux (a magnitude decrease in one or two months) and
no repetition, should be distinguishable from underlying unrelated
sources.

For {\em ASTRO-H}, we have performed simulations for the designed
detector's response by using some of the synthetic spectra as an
input. We have used the detector's response and background presented by \citet{tajima2010} and  \citet{takahashi2010}, and this corresponds to the `optimistic' case in Table 1. 
Figure 7 shows examples of the simulations for the three models (DD2D\_asym\_04, W7, DD2D\_iso\_04) at a distance of 5, 15,
and 25 Mpc. We have used synthetic spectra at 20 days after the explosion. This is indeed conservative, since only the brightest model (DD2D\_asym\_04) reaches the peak at 20 days, and the other two models have not yet reached the maximum flux at 20 days. The simulation assumes an exposure of $10^{6}$ seconds. For the spectral region around the 158 keV line,
all these three models are detectable at 25 Mpc, confirming our estimate in Table 1 (the simulation indeed indicates that values in Table 1 are even underestimated). 
Detecting the signal as a `line' is more difficult -- for 
example, at 15Mpc we can still identify the 158 keV `line' in the
spectrum of the DD2D\_asym\_04 model, but not for the other two
models. The simulation also shows that resolving the 158 keV line
profiles can be definitely done for SNe Ia at 5 Mpc, but it is not
possible for SNe Ia at a distance of 10 Mpc or larger. A slightly
worse level of agreement between the simple estimate and the response
simulations is found for the HXI.
 
\section{Conclusions and Discussion}
In this paper, we have reported properties of high energy emissions
from the radioactive decay chain $^{56}$Ni $\to$ $^{56}$Co $\to$
$^{56}$Fe in SNe Ia. A series of two-dimensional delayed-detonation
models have been investigated.

We estimate, for the narrow-line sensitivity of $3 \times 10^{5}$ photons cm$^{-1}$ s$^{-1}$, 
that the 847
keV line from the decay of $^{56}$Co is detectable by 
SPI/{\em INTEGRAL} for the most-nearby SNe within $\sim$6 Mpc, at 60 days
after the explosion and thereafter. 
This, however, is likely optimistic in view of the non-detection of the signal 
from SN 2011fe by SPI (\S 3.3).

The flux of the 847 keV line is sensitive to $M$($^{56}$Ni)
but not to other model details, such as the progenitor mass,
the flame propagation modes, 1D or multi-D, or the viewing direction.  Thus, the
upper limit of the SPI observation is directly translated to a
constraint on the mass of $^{56}$Ni, as $M$($^{56}$Ni) $\lsim 1.0$ \msun. This is not as strong as the constraint placed by optical
emission analysis \citep{nugent2011b,roepke2012}, but is totally
independent from and more direct than the optical emission
analysis. This shows a potential to place a strong constraint on the
nature of the explosion through the high energy emission.  In the
earlier phase the most constraining signal is the 158 keV line. The
behavior here is sensitive to different models (e.g., the
thermonuclear flame modes, initial conditions and viewing angles
within the delayed-detonation scenario) -- the feature essentially
traces how much material is present atop of the $^{56}$Ni-rich
region.  SN 2011fe was observed by {\em INTEGRAL} with an exposure of
$\sim$$10^6$ seconds starting at $\sim$5 days after the discovery
\citep{isern2011a}, but unfortunately the reported upper limit is not
deep enough to reject any models presented in this paper (including
the pure-detonation model). For more detailed and quantitative
analysis, variations in the line width predicted for different models
will need to be taken into account.

While most previous studies focused on the detectability of the
radioactive signals in the MeV range, we suggest that detecting soft
$\gamma$-rays and hard X-rays is more promising with the new,
near-future observatories ({\em Astro-H} and {\em NuStar}).  We have
found that the 158 keV line is detectable up to $\sim$25 Mpc with SGD
on board {\em ASTRO-H}, and the hard X-ray continuum up to $\sim$15 Mpc with HXI on board {\em ASTRO-H} and {\em NuStar}.  These
near-future observatories, which we predict are able to detect the
high energy emission almost annually, are expected to provide
practically applicable diagnostics on the explosion mechanism.
\begin{itemize}
\item The hard X-ray continuum provides a measure of the composition
  near the surface. 
\item The 158 keV line flux provides a measure of how much material is
  present above the $^{56}$Ni-rich region. 
\item Accordingly, some line-to-continuum ratios, shown to be
  accessible by these new instruments, will provide a strong
  constraint on the explosion mechanism (or the viewing direction in
  the 2D delayed-detonation models).
\end{itemize} 

Compared to other model variants 
(e.g., 1D pure deflagration model, 1D pure detonation model), 
the 2D delayed detonation model tends to predict a
lower (angle-averaged) flux in these energy ranges. Thus our
estimate on the detectability of these features with the future
observatories may well be a conservative estimate. The hard X-ray
continuum and the 158 keV line data alone could be used to obtain a
rough constraint on $M$($^{56}$Ni), but this is contaminated by the
factors arising from different explosion models as described above.

According to the systematic study of the 2D delayed-detonation models,
we have the following solid prediction for this model sequence: For a
large statistical sample, a possible scatter in the peak flux may
limit the degree of the asymmetry in the explosions. We predict larger
asymmetry, thus a larger scatter in the peak flux, for fainter SNe Ia
according to the delayed-detonation model.

Future new generation $\gamma$-ray observatories like {\em GRIPS} are
expected to deepen the observable horizon in the MeV range up to $\sim$20 Mpc, or even to 35 Mpc for an exposure as long as $3 \times
10^{6}$ seconds. The 812 keV line can be used in a way similar to the
emission features in the softer band as described above. 
The peak 847 keV line flux alone is
a very good tracer for $M$($^{56}$Ni); for the model sequences we explore in the paper (the 2D delayed-detonation models, 1D models including various flame propagation modes, those based on Chandrasekhar and sub-Chandrasekhar progenitors), it is insensitive to 
the model variants and the viewing direction, thus being a direct
probe of explosive nucleosynthesis in SNe Ia.

\acknowledgements 
The authors thank Stuart Sim for discussion and useful comments on the
manuscript.  This research is supported by World Premier International
Research Center Initiative (WPI Initiative), MEXT, Japan.
K.M. acknowledges financial support by Grant-in-Aid for Scientific
Research for Young Scientists (23740141) and by the Max-Plank Society
as a short-term visitor.  The work of F.R. was supported by the Emmy
Noether Program (RO~3676/1-1) of the Deutsche Forschungs\-gemeinschaft
and the ARCHES award.

\begin{figure*}
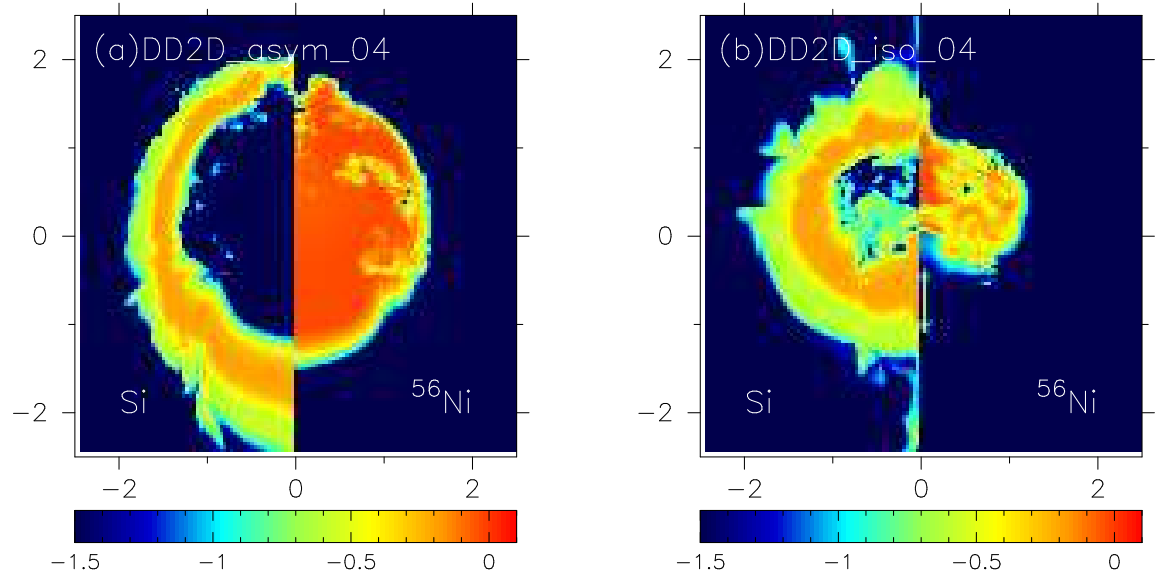

\begin{center}
        \begin{minipage}[]{0.45\textwidth}
                \epsscale{1.0}
                 \plotone{f1a.eps}
        \end{minipage}
       \begin{minipage}[]{0.45\textwidth}
                \epsscale{1.0}
                 \plotone{f1b.eps}
        \end{minipage}
\end{center}
\caption{Examples of the 2D delayed-detonation models taken from \citet{kasen2009}. 
In each panel, the mass fractions of Si (left) and $^{56}$Ni (right) are shown, on a logarithmic scale. The axes are in $10,000$ km s$^{-1}$. 
\label{fig1}}
\end{figure*}

\begin{figure*}
\begin{center}
        \begin{minipage}[]{0.95\textwidth}
                \epsscale{1.0}
                 \plotone{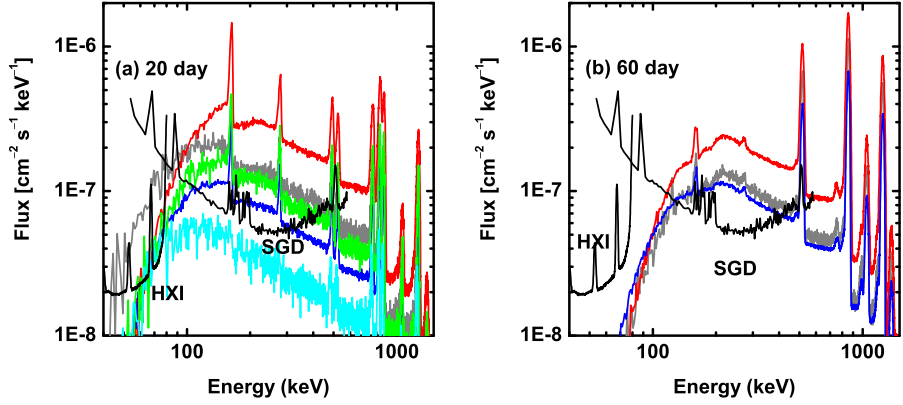}
        \end{minipage}
\end{center}
\caption{Examples of synthetic spectra at (a) 20 days and (b) 60 days after the explosion.  Shown here are angle-averaged spectra for models DD2D\_asym\_04 (dc2; red line), W7 (gray), and DD2D\_iso\_04 (dc3; dark blue).  The masses of $^{56}$Ni are 1.02 \msun, 0.64 \msun, and 0.42 \msun, respectively (Tab. 1).  The distance is assumed to be 10 Mpc.  The angle-dependent spectra seen from two opposite directions are shown for DD2D\_iso\_04 (green and cyan) at 20 days. At 60 days the angle dependence is small and thus not shown here. The angle dependence is small for DD2D\_asym\_04 at both epochs. Sensitivity curves for an exposure with $10^6$ seconds of {\em HXI} and {\em SGD} on board {\em Astro-H} \citep[as presented in][]{tajima2010,takahashi2010} are shown by black lines.
\label{fig2}}
\end{figure*}

\clearpage
\begin{figure*}
\begin{center}
        \begin{minipage}[]{0.95\textwidth}
                \epsscale{0.9}
                 \plotone{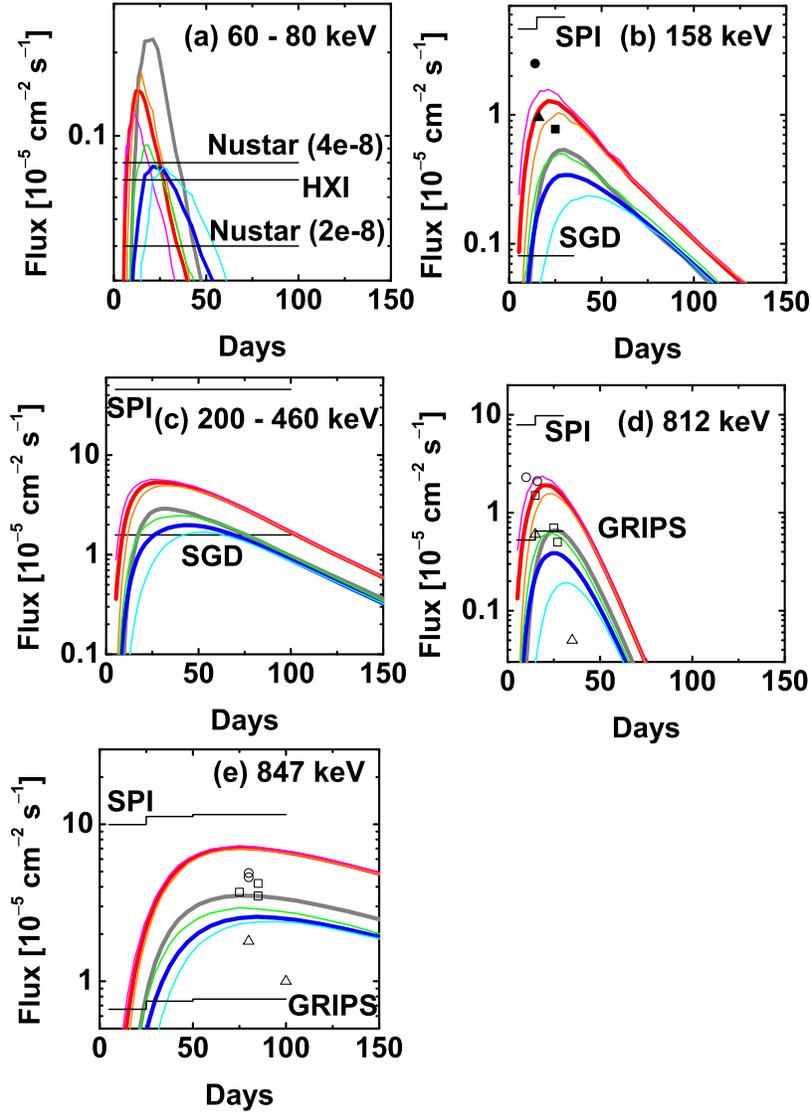}
        \end{minipage}
\end{center}
\caption {Examples of synthetic light curves for (a) hard X-ray continuum, (b) 158 keV line, (c) integrated flux in $200 - 460$ keV, (d) 812 keV line, and (e) 847 keV line.  The distance is assumed to be 10 Mpc, and the flux is normalized at $10^{-5}$ cm$^{-2}$ s$^{-1}$.  Shown here are light curves for W7 (thick gray), angle-averaged (thick red) and angle-dependent (thin magenta and orange) light curves for DD2D\_asym\_04 (dc2), angle-averaged (thick blue) and angle-dependent (thin green and cyan) light curves for DD2D\_iso\_04 (dc3).  For the angle-dependent curves, two opposite directions are shown.  For comparison, we show the predictions on the peak dates and peak fluxes for other model variants, for the 158 keV flux from \citet{gomez1998} and for the 812 and 847 keV fluxes from \citet{milne2004}. For the 158 keV case, the following three models are shown: A pure detonation model DET (filled circle), a delayed-detonation model DEL (filled square), and a sub-Chandrasekhar model SUB (filled triangle). For the 812 keV and 847 keV cases, the following models are shown: `luminous' SN Ia models HECD and W7DT (open circles), `normal-brightness' models HED8, DD202c, W7 (open squares), and `sub-luminous' models HED6 and PDD54 (open triangles).  Sensitivity curves (with an exposure of $10^6$ seconds) for several current and future instruments are shown by black lines.
\label{fig3}}
\end{figure*}

\clearpage
\begin{figure*}
\begin{center}
        \begin{minipage}[]{0.45\textwidth}
                \epsscale{1.0}
                 \plotone{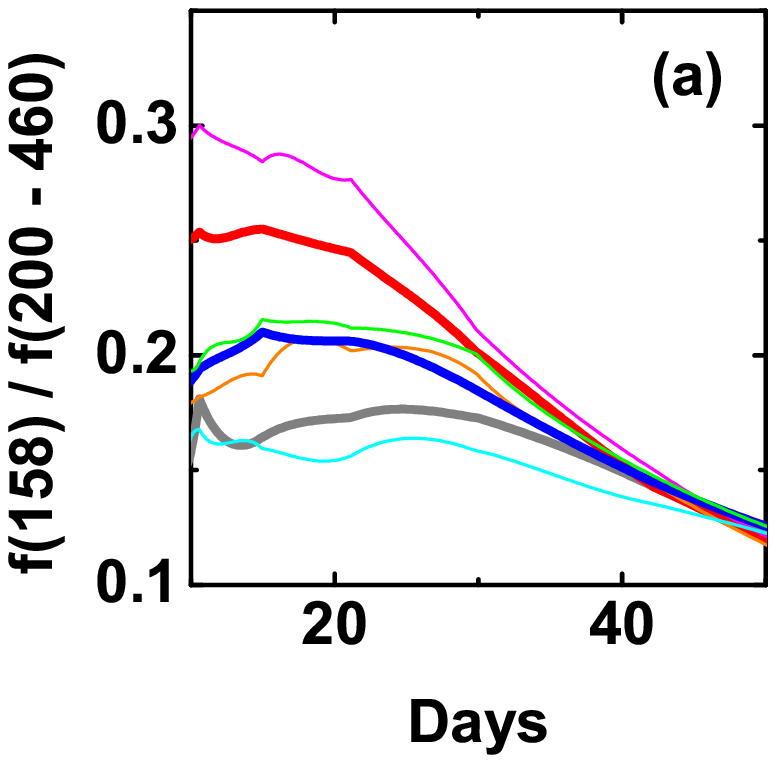}
        \end{minipage}
        \begin{minipage}[]{0.45\textwidth}
                \epsscale{1.0}
                 \plotone{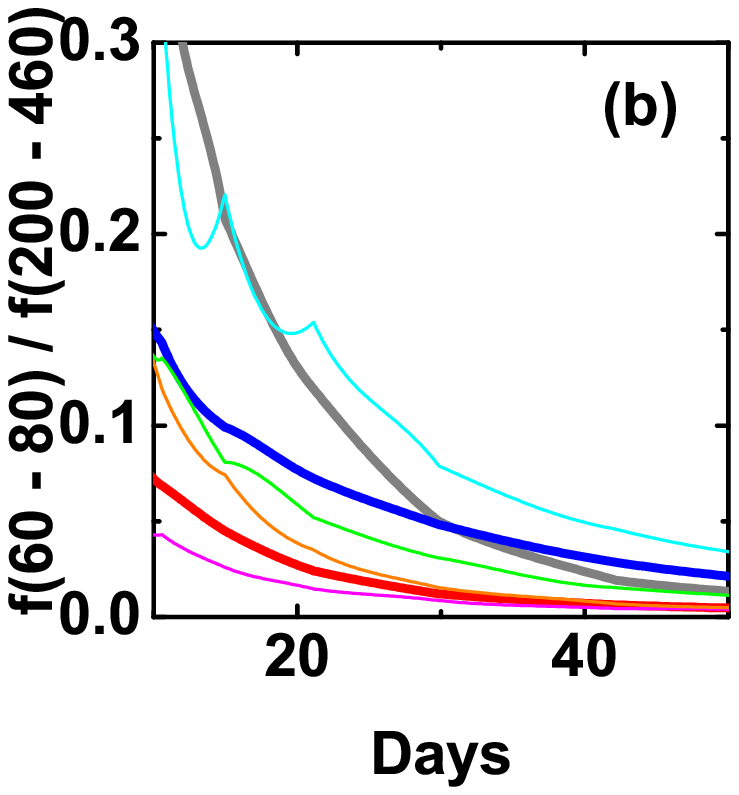}
        \end{minipage}
\end{center}
\caption{Examples of the flux ratio diagnostics, (a) the ratio between the 158 keV line and the $200 - 460$ keV continuum, and (b) the ratio between the $60 - 80$ keV continuum and the $200 - 460$ keV continuum. Shown here are model curves for W7 (thick gray), angle-averaged (thick red) and angle-dependent (thin magenta and orange) curves for DD2D\_asym\_04 (dc2), angle-averaged (thick blue) and angle-dependent (thin green and cyan) curves for DD2D\_iso\_04 (dc3).  For the angle-dependent curves, two opposite directions are shown.  
\label{fig4}}
\end{figure*}

\clearpage
\begin{figure*}
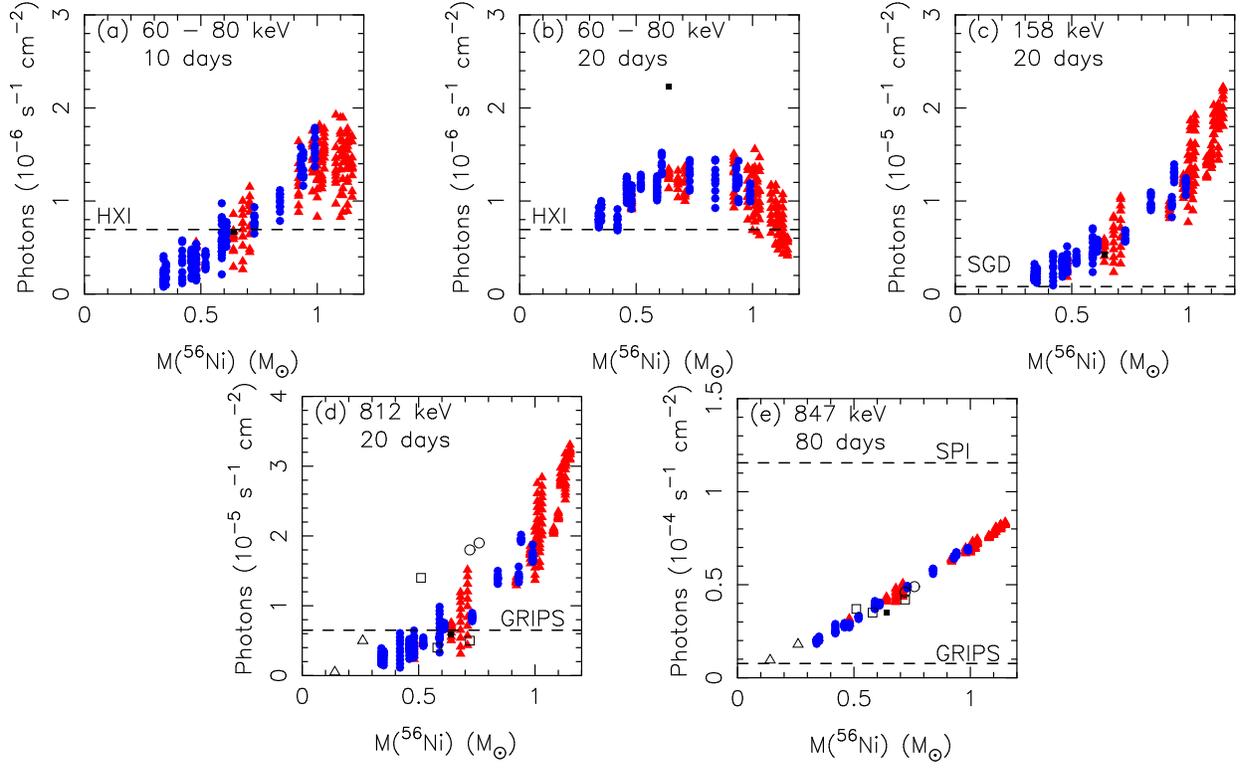

\begin{center}
        \begin{minipage}[]{0.31\textwidth}
                \epsscale{1.0}
                 \plotone{f5a.eps}
        \end{minipage}
        \begin{minipage}[]{0.31\textwidth}
                \epsscale{1.0}
                 \plotone{f5b.eps}
        \end{minipage}
       \begin{minipage}[]{0.31\textwidth}
                \epsscale{1.0}
                 \plotone{f5c.eps}
        \end{minipage}
       \begin{minipage}[]{0.31\textwidth}
                \epsscale{1.0}
                 \plotone{f5d.eps}
        \end{minipage}
       \begin{minipage}[]{0.31\textwidth}
                \epsscale{1.0}
                 \plotone{f5e.eps}
        \end{minipage}
\end{center}
\caption{Selected continuum and line fluxes as a function of $M$($^{56}$Ni) for a distance 
of 10 Mpc. 
(a) 60 -- 80 keV at 10 days, (b) 60 -- 80 keV at 20 days, 
(c) the 158 keV line at 20 days, (d) the 812 keV line at 20 days, and (e) the 
847 keV line at 80 days. In each panel, all 33 models are shown for different viewing 
angles (divided into 10 bins each). 
The `asym' models are shown by red triangles, `iso' models by blue circles, and W7 by 
a black square. For comparison, 1D models from \citet{milne2004} are shown for the 
812 and 847 keV lines (open symbols: See the caption of Figure 3). The sensitivities of various 
current and future instruments are shown for an exposure of $10^6$ seconds. 
\label{fig5}}
\end{figure*}

\clearpage
\begin{figure*}
\begin{center}
        \begin{minipage}[]{0.9\textwidth}
                \epsscale{0.8}
                 \plotone{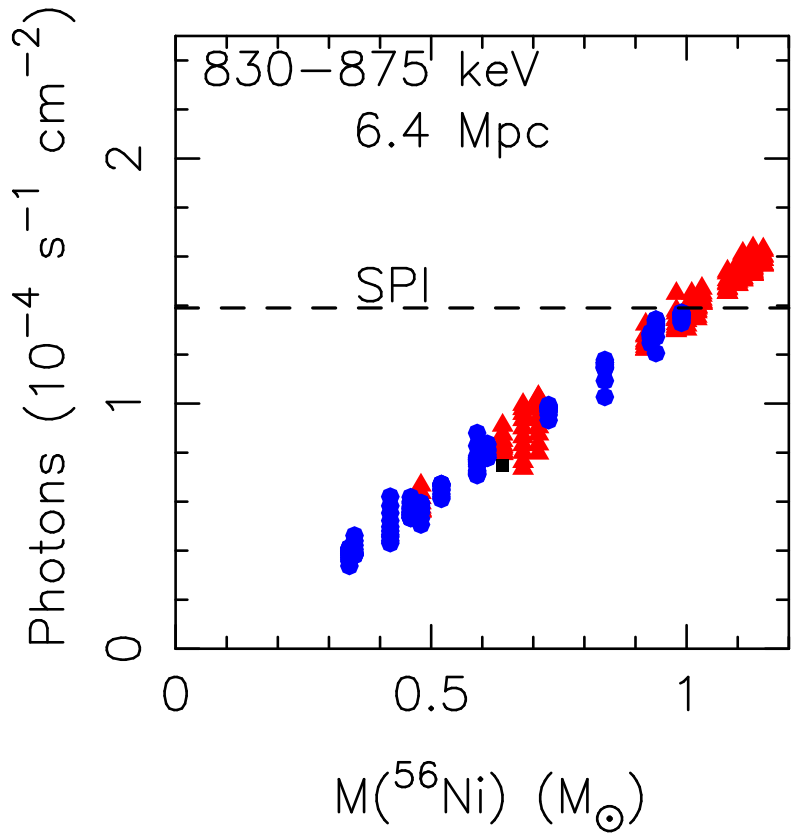}
        \end{minipage}
\end{center}
\caption
{A constraint on the models from the {\em INTEGRAL/SPI} observation of SN 2011fe. The model fluxes for an SN at 6.4 Mpc are time-averaged within days $45 - 88$. The `asym' models are shown by red triangles, `iso' models by blue circles, and W7 by a black square. The 2$\sigma$ upper limit obtained by SPI \citep{isern2011b} is shown by the dashed line. 
\label{fig6}}
\end{figure*}

\clearpage
\begin{figure*}
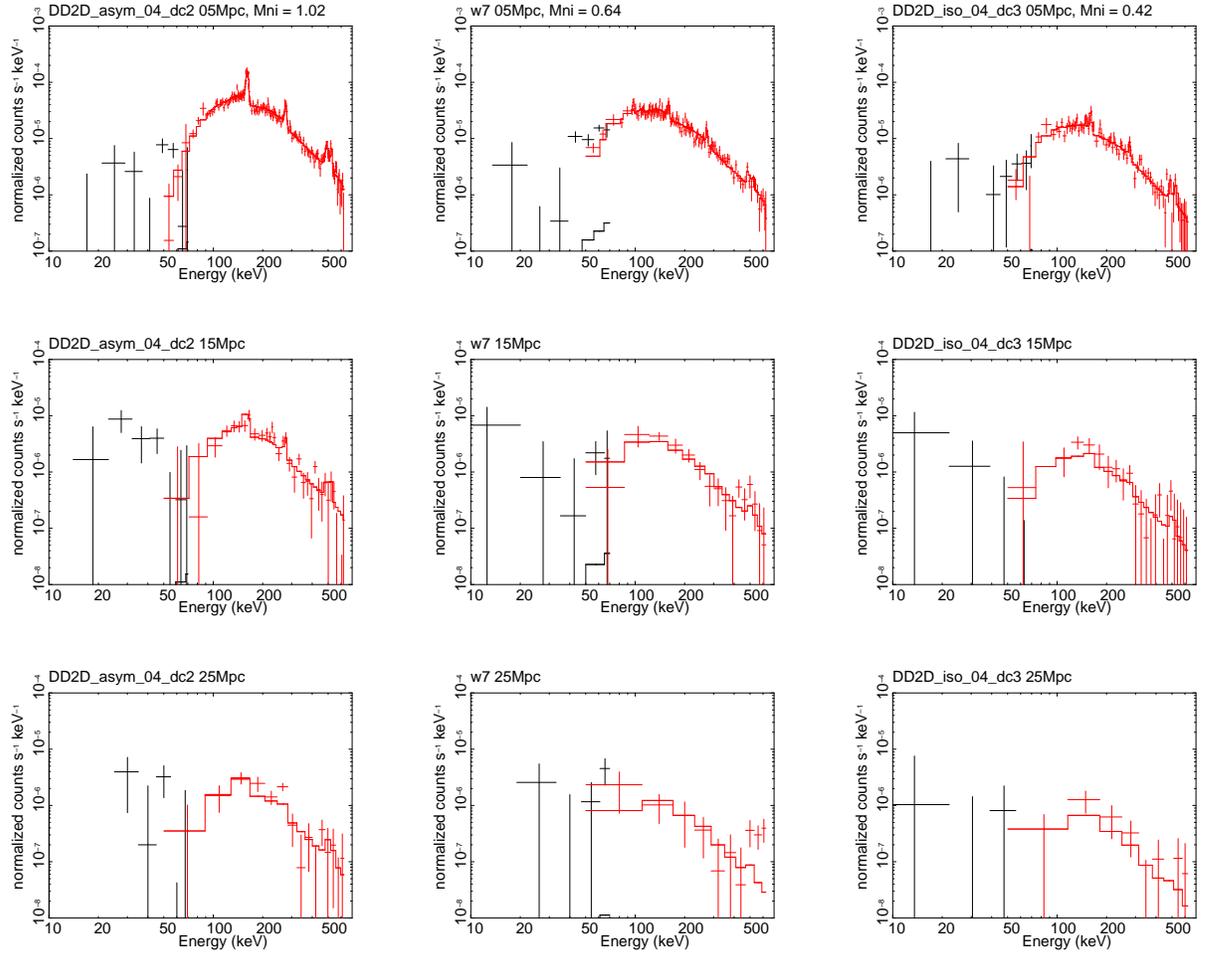

\begin{center}
        \begin{minipage}[]{0.3\textwidth}
                \epsscale{0.75}
                 \plotone{f7a.ps}
        \end{minipage}
        \begin{minipage}[]{0.3\textwidth}
                \epsscale{0.75}
                 \plotone{f7b.ps}
        \end{minipage}
        \begin{minipage}[]{0.3\textwidth}
                \epsscale{0.75}
                 \plotone{f7c.ps}
        \end{minipage}\\
        \begin{minipage}[]{0.3\textwidth}
                \epsscale{0.75}
                 \plotone{f7d.ps}
        \end{minipage}
        \begin{minipage}[]{0.3\textwidth}
                \epsscale{0.75}
                 \plotone{f7e.ps}
        \end{minipage}
        \begin{minipage}[]{0.3\textwidth}
                \epsscale{0.75}
                 \plotone{f7f.ps}
        \end{minipage}\\
        \begin{minipage}[]{0.3\textwidth}
                \epsscale{0.75}
                 \plotone{f7g.ps}
        \end{minipage}
        \begin{minipage}[]{0.3\textwidth}
                \epsscale{0.75}
                 \plotone{f7h.ps}
        \end{minipage}
        \begin{minipage}[]{0.3\textwidth}
                \epsscale{0.75}
                 \plotone{f7i.ps}
        \end{minipage}
\end{center}
\caption
{Detector response simulations for an exposure of $10^{6}$ seconds for some selected models, for HXI (black) and SGD (red) on board {\em Astro-H}. The model spectra at 20 days after the explosion are used as input models, placed at distances of 
5, 15, and 25 Mpc, respectively. We adopt the sensitivity curves from \citet{kokubun2010}, \citet{tajima2010}, and \citet{takahashi2010}. 
\label{fig7}}
\end{figure*}

\clearpage
\begin{deluxetable}{cclll}
 \tabletypesize{\scriptsize}
 \tablecaption{Expected Detectability\tablenotemark{a}
 \label{tab:detectability}}
 \tablewidth{0pt}
 \tablehead{
   \colhead{}
 & \colhead{}
 & \colhead{DD2D\_asym\_04}
 & \colhead{W7}
 & \colhead{DD2D\_iso\_04}
}
\startdata
          & $M$($^{56}$Ni)/\msun  & 1.02 & 0.64 & 0.42\\\hline
Band (keV) & Instrument & Mpc (SNe year$^{-1}$)\tablenotemark{b} & &\\\hline
60--80 & HXI & 13.9 (0.43) & 17.7 (0.96) & 10.5 (0.09)\\
          & NuStar (cons.)\tablenotemark{c} & 13.0 (0.43) & 16.5 (0.70) & 9.7 (0.09)\\
          & NuStar (opt.)\tablenotemark{c} & 18.4 (1.13) & 23.3 (2.52) & 13.8 (0.43)\\
158 & SPI & 4.6 ($<$0.09) & 2.9 ($<$0.09) & 2.3 ($<$0.09)\\
      & SGD (cons.)\tablenotemark{d} & 22.2 (2.09) & 14.2 (0.43) & 11.4 (0.09)\\
      & SGD (opt.)\tablenotemark{d} & 38.5 (6.70) & 24.6 (2.96) & 19.7 (1.57)\\
200--460 & SPI & 3.7 ($<$0.09) & 2.7 ($<$0.09) & 2.3 ($<$0.09) \\
             & SGD (cons.) & 11.6 (0.09) & 8.6 (0.09) & 7.1 (0.09) \\
             & SGD (opt.) & 20.2 (1.74) & 14.8 (0.43) & 12.3 (0.26) \\
812 & SPI & 4.3 ($<$0.09) & 2.6 ($<$0.09) & 2.0 ($<$0.09)\\
      & GRIPS & 16.8 (0.87) & 10.0 (0.09) & 7.6 (0.09)\\
847 & SPI & 7.7 (0.09) & 5.4 ($<$0.09) & 4.6 ($<$0.09)\\
      & GRIPS & 29.8 (4.52) & 21.0 (2.00) & 18.0 (1.04)
\enddata
\tablenotetext{a}{For an exposure of $10^6$ seconds centered at the peak flux.}
\tablenotetext{b}{Limiting distance (expected number of SNe per year), 
computed for the angle-averaged synthetic spectra. The number of SNe is 
estimated from SNe Ia at a redshift below 0.01 (thus roughly complete below 40 Mpc) 
since 2000 taken from the Asiago Supernova Catalog that is constantly updated with new SN discoveries 
\citep{barbon1999}.}
\tablenotetext{c}{Conservative and optimistic estimates assume 
$4$ and $2$ $\times 10^{-8}$ cm$^{-2}$ s$^{-1}$ keV$^{-1}$, respectively, for 
the {\em NuStar} sensitivity \citep{koglin2005}.}
\tablenotetext{d}{The optimistic estimate assumes the designed sensitivity curve of 
SGD as of 2010 \citep{tajima2010,takahashi2010}. Since the instrument design may change, 
we also show a conservative estimate, where we degrade the 2010 sensitivity curve by a 
factor of three.}
\end{deluxetable}

\end{document}